\documentclass[11pt]{article}

\def\no{\, : \,}

\def\kp{\,\dot{+}\,}

\usepackage{amssymb}
\usepackage{citesort}

\begin{document}

\title{%
Scalar field theory on $\kappa$-Minkowski space-time\\
 and Doubly Special Relativity}
\author{M.\ Daszkiewicz
\thanks{email: marcin@ift.uni.wroc.pl},
K.\ Imi\l{}kowska
\thanks{email: kaim@ift.uni.wroc.pl},
 J.\ Kowalski--Glikman
\thanks{email: jurekk@ift.uni.wroc.pl}, and
 S.\ Nowak
\thanks{email: pantera@ift.uni.wroc.pl}\\  \\ {\em Institute for Theoretical
Physics}\\ {\em University of Wroc\l{}aw}\\ {\em Pl.\ Maxa Borna 9}\\
{\em Pl--50-204 Wroc\l{}aw, Poland}} \maketitle

\begin{abstract}
In this paper we recall the construction of scalar field action on $\kappa$-Minkowski space-time and investigate its properties. In particular we show how the co-product of $\kappa$-Poincar\'e algebra of symmetries arises from the analysis of the symmetries of the action, expressed in terms of Fourier transformed fields. We also derive the action on commuting space-time, equivalent to the original one. Adding the self-interaction $\Phi^4$ term we investigate the modified conservation laws. We show that the local interactions on $\kappa$-Minkowski space-time  give rise to 6 inequivalent ways in which energy and momentum can be conserved at  four-point vertex. We discuss the relevance of these results for Doubly Special Relativity.
\end{abstract}
\maketitle

\section{Introduction}
The $\kappa$-Minkowski space \cite{kappaM} is a noncommutative space-time,
in which time and space coordinates do not commute, but instead satisfy the
following Lie-type commutational relation\footnote{Throughout this paper we choose the Planck units, in which the Planck mass scale $\kappa$ as well as the Planck length scale $1/\kappa$ are equal 1. }
\begin{equation}\label{1}
 [t, x_i] = -i x_i.
\end{equation}
It has been  shown in \cite{Kowalski-Glikman:2002jr},
\cite{Kowalski-Glikman:2002ft} that the $\kappa$-Minkowski space-time is a
natural  structure of Doubly Special Relativity (DSR) \cite{Amelino-Camelia:2000ge},
\cite{Amelino-Camelia:2000mn},  \cite{jkgminl},
\cite{rbgacjkg}, a recently formulated theoretical framework aimed to describe
processes that take place at energies close to  Planck energy in the regime,
in which gravity can be neglected. Recently it has been also claimed  that one can interpret DSR as a flat space limit of quantum gravity \cite{Amelino-Camelia:2003xp}, \cite{Freidel:2003sp} or, in other words, quantum
special relativity (see \cite{Kowalski-Glikman:2003hi},
\cite{Amelino-Camelia:2002vy} for recent reviews of the DSR programme.)

The essential physical postulate of the DSR is that there exist in nature two
observer independent scales: of velocity, identified with the velocity of light
 and of mass $\kappa$, which is expected to
be of order of Planck mass. In  Planck units we use in this paper both these scales are equal 1. Since $\kappa$ is to be, by assumption,
observer-independent, in DSR one is forced to modify the way, in which momenta
transform under boost so that this scale is the same for all inertial observers. This is done as follows \cite{jkgminl},
\cite{rbgacjkg}. One assumes that the algebra of
Lorentz transformations, generated by rotations $M_i$ and boosts $N_i$ is not
deformed, i.e.
$$
[M_i, M_j] = i\, \epsilon_{ijk} M_k, \quad [M_i, N_j] = i\, \epsilon_{ijk} N_k,
$$
\begin{equation}\label{2}
  [N_i, N_j] = -i\, \epsilon_{ijk} M_k.
\end{equation}
hold. One assumes also that the action of rotation on momenta is classical,
\begin{equation}\label{3}
  [M_i, k_j] = i\, \epsilon_{ijk} k_k, \quad [M_i, \omega] =0
\end{equation}
however the commutators of boosts and momenta are subject to deformation, to wit\footnote{In this paper we will work in the  bicrossproduct basis of the DSR, called sometimes DSR1. Other bases (DSR
theories) exist \cite{juse}, an example being the Magueijo--Smolin basis (or
DSR2), see \cite{Magueijo:2001cr}. The reason for choosing DSR1 is that this basis is most naturally related with the structure of the $\kappa$-Minkowski space time, and, in particular, with the bicovariant differential calculus we are to use below.}
\begin{equation}\label{4}
   \left[N_{i}, {k}_{j}\right] = i\,  \delta_{ij}
 \left( {1\over 2} \left(
 1 -e^{-2{\omega}}
\right) + {{\mathbf{k}^2}\over 2}  \right) - i\, k_{i}k_{j} ,
\end{equation}
and
\begin{equation}\label{5}
  \left[N_{i},\omega\right] = i\, k_{i}.
\end{equation}
The Casimir of this algebra reads
\begin{equation}\label{6}
 {\cal C}(\omega, \mathbf{k}) = \left(2\sinh{\omega/2}\right)^2 - {\mathbf{k}^2}\, e^{\omega}
\end{equation}
but one should keep in mind that, contrary to Special Relativity, due to the presence of the mass scale, one can construct infinitely many Casimirs with correct low energy ($k, \omega \ll 1$) behavior. Namely any function of ${\cal C}$ will be a Casimir as well, and by appropriate scaling it can always be given dimension of mass square.

One should also notice that by adding the co-products
\begin{equation}\label{copro1}
   \Delta \omega = \omega \otimes 1 + 1 \otimes \omega, \quad \Delta k_i = k_i \otimes 1 + e^{-\omega} \otimes k_i,
\end{equation}
\begin{equation}\label{copro2}
   \Delta M_i = M_i \otimes 1 + 1 \otimes M_i,\quad \Delta N_i = N_i \otimes 1 + e^{-\omega} \otimes N_i + \epsilon_{ijk} k_j \otimes M_k
\end{equation}
antipode, and counit one can extend the algebra (\ref{1})--(\ref{5})  to quantum $\kappa$-Poincar\'e algebra, as originally derived in \cite{qP}.
\newline

In this paper we present some elements of  construction of  field theory on $\kappa$-Minkowski space-time (\ref{1}), which make it possible us to discuss issues relevant for the DSR programme.
There were several past attempts to construct (quantum) field theory on $\kappa$-Minkowski. In
the series of papers \cite{Kosinski:1999ix}, \cite{Kosinski:2001ii}, \cite{Kosinski:2003xx} the authors provide most of the necessary mathematics and essentially constructed  free classical field theory. However as stated in \cite{Kosinski:1999ix}, the formalism seems to leads to
interaction vertices that are not symmetric under relabelling of identical
particles, the feature which is clearly physically unacceptable. Below we show that the interaction is in fact symmetric, as it should be, in full agreement with the earlier results reported in \cite{Amelino-Camelia:2001fd}, where, by using the path integral quantization of a field theory on $\kappa$-Minkowski space-time, it was shown that the vertex in $\Phi^4$ theory is symmetric.  A related approach to construction of QFT on $\kappa$-Minkowski  was presented in \cite{Dimitrijevic:2004nv}.

The plan of the paper is as follows. In the next section, following \cite{Kosinski:1999ix}, \cite{Kosinski:2003xx} we present mathematical
apparatus necessary to formulate QFT on $\kappa$-Minkowski space-time i.e.,
definition of differentiation, integration and Fourier transform. Then, in
Sect.\ III we discuss the variational principle for free scalar field and solutions of field equations. It turns out that the action can be formulated in three equivalent forms: on non-commutative $\kappa$-Minkowski space-time, on the Fourier transformed fields defined as functions of momenta, and finally, on commuting manifold. In Sect.\ IV we briefly discuss a possible quantization of the free scalar field, taking as a starting point the momentum space action and making use of the identification of complex and symplectic structures. Sect.\ V is devoted to discussion of symmetries of the free action in momentum space. We find that the action of symmetries on products of fields must follow the quantum $\kappa$-Poincar\'e co-product rule (\ref{copro1}), (\ref{copro2}), instead of the standard Leibniz one. Analogous result in the case of ($\kappa$-Minkowski) space-time action was derived  earlier in \cite{Agostini:2003vg}. In Sect.\ VI we consider $\Phi^4$ interaction  and we show that it leads to 6 independent energy and momentum conservation laws and briefly discuss  the relevance of this observation in view of DSR programme. Finally, we devote Sect.\ VII to conclusions and discussion of open problems.

\section{Mathematical preliminaries: calculus on $\kappa$-Minkowski space-time}

It is obvious that the calculus is an indispensable tool necessary to construct
classical and quantum field theory. Yet on the non-commutative space-time
differentiation and integration have some unexpected features, which we will
discuss in this section. We follow the presentation
reported in \cite{5dcalc1}, \cite{5dcalc2}, \cite{5dcalc3},
\cite{Kosinski:1999ix}, \cite{Kosinski:2003xx}.

Let us start with the observation that since the $\kappa$-Minkowski coordinates
$t, x_i$ do not commute, one must define an appropriate ordering of functions
defined on this space-time. Here we will adopt the ``time to the left''
ordering, and we will denote such ordering of the function $f(t, x_i)$ by $\no
f(t, x_i)\no$. In particular the ordered on-shell plane wave will have the form
$$
\no \psi_{\mathbf{k}}(t, x)\no = e^{i\omega(\mathbf{k}) t}\, e^{-i \mathbf{k}\,
\mathbf{x}},
$$
where the function $\omega(\mathbf{k})$ is the solution of the mass-shell
condition, as it will be explicitly shown below.

\subsection{Bicovariant differential calculus}

 To define the bicovariant differential calculus (see \cite{5dcalc1},
\cite{5dcalc2}, \cite{5dcalc3}) one considers the total differential of
function defined on $\kappa$-Minkowski space-time. It turns out that
\begin{equation}\label{11}
 d f(t,\mathbf{x}) = \hat\partial_\mu\, f(t,\mathbf{x})\, dx_\mu + \hat\partial_4\, f(t,\mathbf{x})\, dx_4
\end{equation}
and thus the bicovariant calculus is necessarily five
dimensional, and the basic one-forms $dx_\mu, dx_4$ satisfy the following
commutational relations
\begin{equation}\label{12}
  [t,dx_4] = i\, dt, \quad [t,dt] = i\, dx_4, \quad [t,dx_i] = 0,
\end{equation}
\begin{equation}\label{13}
  [x_i, dx_4] = [x_i, dt] =i\, dx_i, \quad [x_i, dx_j] = i
\delta_{ij}(dt - dx_4).
\end{equation}
 It
follows  that the differentials transform in the standard way under action of
rotations and boosts
\begin{equation}\label{15}
 [M_i, dx_j] = i\epsilon_{ijk}\, dx_k, \quad [M_i, dt]=[M_i, dx_4] =0
\end{equation}
\begin{equation}\label{16}
 [N_i,dx_j] = i\, \delta_{ij}\, dt, \quad [N_i, dt] = i dx_i,\quad [N_i, dx_4] =0
\end{equation}
This fact can be easily understood in the de Sitter geometric framework of DSR
theories, see \cite{Kowalski-Glikman:2003we}.

With the help of
Leibniz rule, $d(fg) = df\, g + f\, dg$ one can check that for any function
$\phi(t,\mathbf{x})$
\begin{equation}\label{17}
 \hat\partial_A \phi(t,\mathbf{x}) =  \no
\eta_A\left(\frac1i\frac{\partial}{\partial t},
 -\frac1i\frac{\partial}{\partial \mathbf{x}}\right)\,\phi(t,\mathbf{x}) \no,
\end{equation}
where  $\eta_A(\omega, \mathbf{k})$ are given by
\begin{eqnarray}
 {\eta_0}(\omega, \mathbf{k}) &=&  \sinh
{\omega} + \frac{\mathbf{k}^2}{2}\,
e^{  {\omega}} \nonumber\\
 \eta_i(\omega, \mathbf{k}) &=&   k_i \, e^{  {\omega}} \nonumber\\
 {\eta_4}(\omega, \mathbf{k}) &=&
  \cosh {\omega}  - \frac{\mathbf{k}^2}{2} \,
e^{ {\omega}}-1.   \label{14}
\end{eqnarray}
In particular, for the ordered plane wave (off-shell) we find
\begin{equation}\label{18}
\hat\partial_0 \, e^{i\omega t}\, e^{-i \mathbf{k}\,
\mathbf{x}} =  \left( \sinh {\omega} +
\frac{\mathbf{k}^2}{2}\, e^{  {\omega}}\right)\,e^{i\omega
t}\, e^{-i \mathbf{k}\, \mathbf{x}}
\end{equation}
\begin{equation}\label{19}
\hat\partial_i \, e^{i\omega t}\, e^{-i \mathbf{k}\,
\mathbf{x}} =   \left(k_i\, e^{\omega }\right)\, e^{i\omega t}\, e^{-i
\mathbf{k}\, \mathbf{x}}
\end{equation}
One can also use the operator $\hat\partial\equiv \hat\partial_4$
\begin{equation}\label{19a}
    \hat\partial \, e^{i\omega t}\, e^{-i \mathbf{k}\,
\mathbf{x}} =  \left( \cosh {\omega} -
\frac{\mathbf{k}^2}{2}\, e^{  {\omega}}-1\right)\,e^{i\omega
t}\, e^{-i \mathbf{k}\, \mathbf{x}} =  \frac12\, {\cal C}(\omega, \mathbf{k}) \,e^{i\omega
t}\, e^{-i \mathbf{k}\, \mathbf{x}}
\end{equation}
(cf.\ (\ref{6}).) Note that it follows implicitly from (\ref{11}) and explicitly from (\ref{17}),
(\ref{14}) that since the differentials $dx_\mu$ transform in the standard way
under action of Lorentz generators, the derivatives $\hat\partial_\mu$ are the
standard Lorentz vectors as well. Thus the second order operator
$\left(\partial_0^2 -\partial^2_i\right)$, as well as the first order operator $2\hat\partial$ are Lorentz scalars with correct low energy behavior and can be
employed as the natural kinetic operator for a scalar field.

\subsection{Integration and Fourier transform}

Let us now turn to the definition  of integral over the $\kappa$-Minkowski
space-time. This integral is defined for ordered functions
$$
\no f(t,\mathbf{x}) \no = \sum f^{(1)}(t)f^{(2)}(\mathbf{x})
$$
and reads\footnote{To avoid ambiguities, arising as a result of the nontrivial commutators between differentials and positions (\ref{12}), (\ref{13}), we always write the measure to the left of the integrand.}
\begin{equation}\label{22}
\int \, d^4x \no f(t,\mathbf{x}) \no
\end{equation}
Note that since  the
differentials $dx_\mu$ transform in the standard way under Lorentz
transformation, the integral of a Lorentz scalar is Lorentz invariant.

Having the integral, we can introduce the delta function to be ($ikx \equiv  i\omega t - i \mathbf{k}\mathbf{x}$)
\begin{equation}\label{23}
 \delta(k) = \frac{1}{(2\pi)^4}\,   \int \, d^4x \no e^{ikx} \no \left( \equiv \frac{1}{(2\pi)^4}\,   \int \, dt\, e^{i\omega t}\, \int \, d^3\mathbf{x}\,  e^{- i \mathbf{k}\mathbf{x}}\right)
\end{equation}
We can also introduce a Fourier transformed function $\tilde{\Phi}(k)$ by
\begin{equation}\label{24}
    \Phi(t, \mathbf{x}) = \frac{1}{(2\pi)^4}\,   \int \, d\mu\, \tilde{\Phi}(\omega, \mathbf{k}) \, e^{i\omega t}\, e^{- i \mathbf{k}\mathbf{x}}
\end{equation}
where $d\mu \equiv d\omega d^3\mathbf{k}\, e^{3\omega}$ is the
measure on the momentum space, which, thanks to the $e^{3\omega}$
factor is Lorentz invariant. From these formulae one finds
\begin{eqnarray}
\int \, d^4x \Psi(x) \Phi(x) &=& \int  d^4x \int d\mu d\mu'\,   \,
\tilde{\Psi}(k) \tilde{\Phi}(k')\no e^{ikx} \no  \no
e^{ik'x} \no\nonumber\\
\label{25} &=& \int\, d\mu\,  d\mu' \, \tilde{\Psi}(k)\,
\tilde{\Phi}(k')\, \delta(k \kp k')
\end{eqnarray}
To derive eq.\ (\ref{25}) one uses the fact that
\begin{equation}\label{26}
    \no e^{ikx} \no \no e^{ik'x} \no = \no e^{i(k \kp k')x} \no
\end{equation}
where, as a result of the non-commutative structure of $\kappa$-Minkowski space-time (\ref{1}),
\begin{equation}\label{27}
    k \kp k' = (\omega+\omega'; e^{-\omega'} \,\mathbf{k} +  \mathbf{k'})
\end{equation}

\section{Actions and field equations}

In view of the differential calculus described in the preceding section there are two potential expressions for the lagrangian: $$
L=\left(\Phi(x)\, \hat{\partial} \, \Phi(x) - M^2\, \Phi(x)^2\right),
$$
and
$$
L'=\frac12\left(\eta^{\mu\nu} \, \Phi(x)\, \hat{\partial}_\mu \hat{\partial}_\nu\, \Phi(x) - M^2\, \Phi(x)^2\right).
$$

In the following we choose the first one for its simplicity, but the difference is only in the form of the function ${\cal M}(\omega,\mathbf{k})$ (see \ref{29} below) that leads to the mass-shell condition. Thus our action reads
\begin{equation}\label{28}
S = \int\, d^4x \, L
\end{equation}
In terms of the Fourier transformed fields (\ref{24}) this action reads (from now on we denote the Fourier transformed fields without tilde)
\begin{equation}\label{29a}
  S = \frac12 \int\, d\omega d^3 \mathbf{k}\,  {\Phi}(\omega, \mathbf{k})\, {\Phi}(-\omega, - e^{\omega} \,\mathbf{k})\, \left[ \left(2\sinh{\omega/2}\right)^2 - {\mathbf{k}^2}\, e^{\omega}   - M^2\right]
\end{equation}
while
\begin{equation}\label{29}
  S' = \frac12 \int\, d\omega d^3 \mathbf{k}\,  {\Phi}(\omega, \mathbf{k})\, {\Phi}(-\omega, - e^{\omega} \,\mathbf{k})\, \left[ \sinh^2\omega -  \frac12\, \mathbf{k}^2\left( e^{2\omega} +1\right) + \frac{\mathbf{k}^4}{4}\, e^{2\omega} - M^2\right]
\end{equation}
Note that  the factors $e^{3\omega}$ in the integration measure cancel. This action can be also expressed in slightly more compact way using the antipode $S$ (generalized ``minus'') of the $\kappa$-Poincar\'e algebra
$$
S(\omega)=-\omega, \quad S(\mathbf{k}) = - e^\omega\, \mathbf{k}
$$
as
\begin{equation}\label{29b}
  S = \frac12 \int\, d\omega d^3 \mathbf{k}\,  {\Phi}(\omega, \mathbf{k})\, {\Phi}(S(\omega), S(\mathbf{k}))\, {\cal M}(\omega,\mathbf{k})
\end{equation}
where ${\cal M}(\omega,\mathbf{k})$ denotes the term in the
parentheses in (\ref{29a}) or (\ref{29}) or any other Lorentz
invariant function, since, as mentioned above, there is infinitely
many of a priori equally good Casimirs for $\kappa$-Poincar\'e
algebra. The action (\ref{29b}) will be the basis of our
investigations below.

Varying this action with respect
to ${\Phi}(\omega, \mathbf{k})$ and noting that ${\cal M}(\omega,\mathbf{k}) = {\cal M}(S(\omega),S(\mathbf{k}))$   we find the on-shell condition of the form
\begin{equation}\label{30}
 {\cal M}(\omega,\mathbf{k}) =    \left(2\sinh{\omega/2}\right)^2 - {\mathbf{k}^2}\, e^{\omega}  - M^2 =0.
\end{equation}
Using (\ref{30}), we can write down the on-shell field (operator)
in the usual way as follows
\begin{equation}\label{on-shell}
   \Phi(t,\mathbf{x}) = \int d\omega d^3 \mathbf{k}\, \, e^{3\omega}\,\Phi(\omega, \mathbf{k})\, e^{i \omega t} e^{-i\mathbf{k}\mathbf{x}} \, \delta\left({\cal M}(\omega, \mathbf{k})\right)
\end{equation}

In \cite{Kosinski:1999ix} the authors noticed that it might be possible to find an action equivalent to (\ref{28}) (\ref{29}) by making the inverse Fourier transform from ${\Phi}(\omega, \mathbf{k})$ to position space fields, defined on commuting space-time. Let us therefore turn to the construction of such action. Let us denote the commuting time  by $\tau$ and commuting space by $\mathbf{y}$. Then
\begin{equation}\label{ca1}
   {\Phi}(\omega, \mathbf{k}) = \int d\tau\, d^3 \mathbf{y}\, \Phi(\tau, \mathbf{y})\, e^{i\omega\tau - i \mathbf{k}\mathbf{y}}
\end{equation}
Substituting this into the lagrangian (\ref{29}), we get
\begin{eqnarray}
\hspace{-15pt}S &=& \frac12 \int\, d\omega d^3 \mathbf{k}\, \int\,
d\tau d^3 \mathbf{y}\, d\tau' d^3 \mathbf{y}'\, \Phi(\tau,
\mathbf{y})\, \Phi(\tau', \mathbf{y}')\, {\cal M}\left(\frac1i\,
\frac{\partial}{\partial \tau}; -\frac1i\,
\frac{\partial}{\partial\mathbf{y}}\right) e^{i\omega\tau - i
\mathbf{k}\mathbf{y}}\, e^{-i\omega\tau' + i e^\omega\,
\mathbf{k}\mathbf{y}'} \nonumber \\
&=&\frac12 \int\, d\omega \, \int\, d\tau d^3 \mathbf{y}\, d\tau'
d^3 \mathbf{y}'\,\left[ {\cal M}\left(-\frac1i\,
\frac{\partial}{\partial \tau}; \frac1i\,
\frac{\partial}{\partial\mathbf{y}}\right)\Phi(\tau,
\mathbf{y})\right]\, \Phi(\tau', \mathbf{y}')\,  e^{i\omega(\tau -
\tau')} \, \delta(\mathbf{y} -  e^\omega\, \mathbf{y}') \nonumber
\\
&=& \frac12 \int_0^\infty\, dz \, \int\, d\tau d\tau' d^3
    \mathbf{y}\,
     \left[ {\cal M}\left(-\frac1i\, \frac{\partial}{\partial \tau}; \frac1i\,
     \frac{\partial}{\partial\mathbf{y}}\right)\Phi(\tau, \mathbf{y})\right]\,
     \Phi(\tau', z\, \mathbf{y})\,  z^{2-i(\tau - \tau')}\label{ca2}
\end{eqnarray}

It should be noticed that this action leads to much more complex equations of motions
than the ones used to analyze the ``trans-Planckian problem'' in cosmology and black hole
physics in the framework of naive $\kappa$-Poincar\'e algebra \cite{Blaut:2001fy}.
We will investigate the properties of the action (\ref{ca2}) in a separate paper.

\section{Complex structure and canonical commutational relations}

In the preceding section we formulated three equivalent action principles: on $\kappa$-Minkowski space (\ref{28}), on momentum space (\ref{29}), and on the associated commutative space (\ref{ca2}). Of these action the one defined on momentum space is clearly the simplest. However, to proceed with this action, we must reformulate somehow the standard procedure of field theory quantization, which usually relies heavily on position space concepts. For example to find out what the canonical commutational relations (CCR) are one usually identifies field momenta as time derivatives of the fields, and then uses the CCRs of fields and field momenta to derive the commutators of creation and annihilation operators. But to start this procedure one must identify time variable, which is clearly not present in the momentum space picture, and to proceed we clearly need another equivalent set of quantization principles.

In the standard quantum field theory using the canonical symplectic structure on the space of (on-shell) fields and field momenta (i.e.\ assuming that fields and their momenta commute to $\delta$ function) one {\em derives} the commutational relation between creation and annihilation operators of the form
$$
[a_{\mathbf{k}}, a^\dag_{\mathbf{k}'}] = \delta(\mathbf{k}-\mathbf{k}')
$$
where $a^\dag$ is an operator adjoint to $a$, and the creation and annihilation operators are proportional to positive energy, on-shell, momentum space fields, and their adjoint, respectively. Thus we see that in the standard case there is  one-to-one correspondence between the canonical commutational relations and complex structure.

Here, following \cite{Rovelli:1999fq} we take this observation as a starting point of the quantization procedure. We assume that the symplectic structure on the  momentum space fields $\Phi(\omega, \mathbf{k})$ is identified with the complex structure associated with these fields. Consider our action
\begin{equation}\label{IV.1}
  S = \frac12 \int\, d\omega d^3 \mathbf{k}\,  {\Phi}(\omega, \mathbf{k})\, {\Phi}(S(\omega), S(\mathbf{k}))\, {\cal M}(\omega,\mathbf{k})
\end{equation}
Defining the complex conjugate field as
\begin{equation}\label{IV.2}
   {\Phi}^*(\omega, \mathbf{k}) = e^{a\omega}\, {\Phi}(S(\omega), S(\mathbf{k}))
\end{equation}
we see that the action is real. The coefficient $a$ can be fixed
by assuming that the field (\ref{24}) is real as well.

To check reality  of ${\Phi}(t, \mathbf{x})$, we must first decide how the plane waves change under complex conjugation. It is most natural to assume that plane waves are pure phase. Then the complex conjugation has to be defined as
\begin{equation}\label{IV.3}
 \left( e^{i\omega t} e^{-i\mathbf{k}\mathbf{x}}\right)^* = e^{i\mathbf{k}\mathbf{x}} e^{-i\omega t} = e^{-i\omega t}e^{ie^\omega\, \mathbf{k}\mathbf{x}}
\end{equation}
since it leads to
$$
\left( e^{i\omega t} e^{-i\mathbf{k}\mathbf{x}}\right)^*\, e^{i\omega t} e^{-i\mathbf{k}\mathbf{x}} = e^{i\omega t} e^{-i\mathbf{k}\mathbf{x}}\,\left( e^{i\omega t} e^{-i\mathbf{k}\mathbf{x}}\right)^* =1.
$$
Now
\begin{eqnarray}
   {\Phi}^*(t, \mathbf{x}) &=& \frac{1}{(2\pi)^4}\,
   \int \, d\omega d^3 \mathbf{k}\, e^{3\omega}\,
   {\Phi}^*(\omega,  \mathbf{k}) \, e^{-i\omega t}e^{ i\, e^\omega\, \mathbf{k}\mathbf{x}}
   \nonumber \\
&=& \frac{1}{(2\pi)^4}\,
 \int \, d\omega d^3 \mathbf{k}\, e^{-a\omega}\,
 {\Phi}(\omega,  \mathbf{k}) \, e^{i\omega t}e^{- i \mathbf{k}\mathbf{x}}\label{IV.4}
\end{eqnarray}
which is equal to ${\Phi}(t, \mathbf{x})$ if $a =-3$.

Knowing the complex structure on the space of fields, we can now postulate that it is equal to the symplectic structure on this space. Using this postulate we find the canonical commutational relations of the form
\begin{eqnarray}
\left[{\Phi}(\omega,  \mathbf{k}), {\Phi}^*\left(\omega',
\mathbf{k}'\right)\right] &=& \delta(\omega-\omega')\,
\delta(\mathbf{k}-\mathbf{k}'), \nonumber \\
   \left[{\Phi}(\omega,  \mathbf{k}), {\Phi}\left(S(\omega'),  S(\mathbf{k}')\right)\right]
   &=& e^{3\omega}\, \delta(\omega-\omega')\, \delta(\mathbf{k}-\mathbf{k}')\label{IV.5}
\end{eqnarray}
Having these relations one is tempted to proceed to the construction of Fock space. For creation and annihilation operators we could choose
\begin{equation}\label{IV.6}
   \hat a_{\mathbf{k}} =  {\Phi}(\omega(\mathbf{k}),  \mathbf{k}), \quad \hat a^\dag_{\mathbf{k}} =  {\Phi}^*(\omega(\mathbf{k}),  \mathbf{k}) = e^{-3 \omega(\mathbf{k})}\, {\Phi}\left(S(\omega(\mathbf{k})),  S(\mathbf{k})\right),
\end{equation}
where  by using the symbol $\omega(\mathbf{k})\geq0$ we indicate that the fields are positive frequency and on-shell (cf.\ (\ref{30})). It is then possible to define the vacuum as a state annihilated by all the annihilation operators $\hat a_{\mathbf{k}}\, |0>=0$ and to define one-particle DSR state with momentum ${\mathbf{k}}$ as $\hat a^\dag_{\mathbf{k}}\, |0>$. Then the two-particle state would be defined as
\begin{equation}\label{IV.7}
    \hat a^\dag_{\mathbf{k}}\,\hat a^\dag_{\mathbf{k}'}\, |0>
\end{equation}
 This last definition, however, poses a problem. Namely, as we will show in the next section it seems impossible to associate energy and momentum with the state of the form (\ref{IV.7}), at least within the framework of mathematics we have developed so far, the reason being that the function $\hat a^\dag_{\mathbf{k}}\,\hat a^\dag_{\mathbf{k}'}$ cannot be expressed as a result of integration of any local function on $\kappa$-Minkowski space-time. It remains an open question if this problem could be overcome by extending the scope of mathematical tools, or rather by revising the construction of the multi-particles DSR states.

\section{Quantum $\kappa$-Poincar\'e symmetry}

Let us now turn to checking explicitly that the action (\ref{29}) is indeed invariant under deformed Poincar\'e transformation. The invariance of the position space action has been explicitly checked already in \cite{Kosinski:1999ix}. An extremely clear derivation and discussion can be found in \cite{Agostini:2003vg}, where the role of coproduct was emphasized. However, the momentum space action (\ref{29}) is defined on commutative space, and it seems at first sight unclear how the coproduct could enter the structure of symmetries in this case. Below we find that the use of the coproduct rule for action of symmetries on product of functions arises naturally also in momentum space. 

Let us start with deformed boosts (\ref{4}), (\ref{5})
\begin{equation}\label{V.1}
    \delta^B\omega=\varepsilon^i\, k_i, \quad \delta^B k_i = \varepsilon_i \left( {1\over 2} \left(
 1 -e^{-2{\omega}}
\right) + {{\mathbf{k}^2}\over 2}  \right) - \varepsilon^j\, k_{i}k_{j}
\end{equation}
As we will see in a moment, from this procedure, not quite surprisingly, the quantum group co-product structure will emerge. However let us try to proceed naively, to see what goes wrong.

Let us assume that the product  ${\Phi}(\omega, \mathbf{k}){\Phi}(S(\omega), S(\mathbf{k}))$ transforms with respect to (\ref{V.1}) as a scalar, so that,
\begin{eqnarray}
&&\delta^B\left({\Phi}(\omega, \mathbf{k})\, {\Phi}(S(\omega),
S(\mathbf{k}))\right) =\nonumber\\
\hspace{-200pt}&& =\left[\varepsilon^i\, k_i\,
\frac{\partial}{\partial\omega} + \left(\varepsilon_i \left(
{1\over 2} \left(
 1 -e^{-2{\omega}}
\right) + {{\mathbf{k}^2}\over 2}  \right) - \varepsilon^j\,
k_{i}k_{j}\right) \frac{\partial}{\partial k_i}\right]
{\Phi}(\omega, \mathbf{k})\, {\Phi}(S(\omega),
S(\mathbf{k}))\nonumber\\
&& =\left[\frac{\partial}{\partial\omega}\, \varepsilon^i\, k_i +
\frac{\partial}{\partial k_i} \left(\varepsilon_i \left( {1\over
2} \left(
 1 -e^{-2{\omega}}
\right) + {{\mathbf{k}^2}\over 2}  \right) - \varepsilon^j\,
k_{i}k_{j}\right) \right] {\Phi}(\omega, \mathbf{k})\,
{\Phi}(S(\omega), S(\mathbf{k}))\nonumber\\
\label{V.2}
     &&+ 3 \varepsilon^i\, k_i\, {\Phi}(\omega, \mathbf{k})\, {\Phi}(S(\omega), S(\mathbf{k}))
\end{eqnarray}
Since ${\cal M}(\omega,\mathbf{k})$ is invariant under (\ref{V.1})
the transformation of the action reads
\begin{eqnarray}
  \delta^B S &=& \frac12\, \int \delta^B\left(d\omega\, d^3\mathbf{k}\right)\,
  {\cal M}(\omega,\mathbf{k})\,  {\Phi}(\omega, \mathbf{k})\,
  {\Phi}(S(\omega), S(\mathbf{k})) \nonumber \\
\label{V.3}
 &+& \frac12\, \int d\omega\, d^3\mathbf{k}\, {\cal M}(\omega,\mathbf{k})\, \delta^B\left( {\Phi}(\omega, \mathbf{k})\, {\Phi}(S(\omega), S(\mathbf{k}))\right)
\end{eqnarray}

Notice now that since $\delta^B\left(d\omega\, d^3\mathbf{k}\right) = -3 \varepsilon^i\, k_i\, d\omega\, d^3\mathbf{k}$ the last term in (\ref{V.2}) cancels the term resulting from the transformation of the measure. Integrating by parts and observing that
$$
\left[\varepsilon^i\, k_i\, \frac{\partial}{\partial\omega} + \left(\varepsilon_i \left( {1\over 2} \left(
 1 -e^{-2{\omega}}
\right) + {{\mathbf{k}^2}\over 2}  \right) - \varepsilon^j\, k_{i}k_{j}\right) \frac{\partial}{\partial k_i}\right]\, {\cal M}(\omega,\mathbf{k})= 0
$$
we see that  the action (\ref{29}) is indeed invariant under (\ref{V.1}).

We have therefore obtained the desired result,
but it is easy
to observe that our naive procedure
is problematic. To see this, let us return to eq.\ (\ref{V.2}).
It is natural to assume
that the field $\Phi(\omega, \mathbf{k})$ transforms as a scalar as well, i.e.,

\begin{equation}\label{V.2a}
   \delta^B\,{\Phi}(\omega, \mathbf{k}) = \left[\varepsilon^i\, k_i\, \frac{\partial}{\partial\omega} + \left(\varepsilon_i \left( {1\over 2} \left(
 1 -e^{-2{\omega}}
\right) + {{\mathbf{k}^2}\over 2}  \right) - \varepsilon^j\, k_{i}k_{j}\right) \frac{\partial}{\partial k_i}\right] {\Phi}(\omega, \mathbf{k}).
\end{equation}
But then it follows that in order to make
the product
$\Phi(\omega, \mathbf{k})\, \Phi(S(\omega), S(\mathbf{k}))$ transforming as a scalar (\ref{V.2}), the field $\Phi(S(\omega), S(\mathbf{k}))$ should transform exactly as the field $\Phi(\omega, \mathbf{k})$
does, that is, by (\ref{V.2a}). But it follows from (\ref{V.2a}) that we should rather take
\begin{eqnarray}
\delta^B\,{\Phi}(S(\omega), S(\mathbf{k})) &=&
\left[\varepsilon^i\, S(k_i)\, \frac{\partial}{\partial S(\omega)}
+\right.
   \left(\varepsilon_i \left( {1\over 2} \left(
 1 -e^{-2{S(\omega)}}
\right) + {{S(\mathbf{k}^2)}\over 2}  \right) \right. \nonumber \\
&-&\left. \left. \varepsilon^j\, S(k_{i})S(k_{j})\right)
\frac{\partial}{\partial S(k_i)}\right] {\Phi}(S(\omega),
S(\mathbf{k})) \label{V.2b}
\end{eqnarray}
Before solving this puzzle, we will pause for a moment to consider a slightly less complicated case of the symmetries related to space-time translations.
\newline

Let us therefore turn to the remaining part of the (deformed) Poincar\'e symmetry,
namely the symmetries that correspond to translations in  space-time.
At the first sight it seems to be impossible that these symmetries are present in the
momentum space action, as there are no positions, on which the translations could act.
 However, the corresponding symmetries are present. To see this, consider the standard case
 and notice that when space or time coordinate of the field is shifted by $\mathbf{a}$,
  then the Fourier transformed field is multiplied by the phase
  $e^{-i\mathbf{k}\mathbf{a}}$. We see therefore that the  translation in space-time
  fields is in the one-to-one correspondence with the phase transformations of the momentum
  space ones. This suggests that the ten parameter group of Poincar\'e symmetries in
  space-time translates into six parameter Lorentz group plus four independent phase
  transformations in the momentum space, being representations of the same algebra.

Using this insight let us  turn to the case at hands. Consider first time translations. The corresponding infinitesimal phase transformation reads\footnote{Note that since the function ${\cal M}$ is real, $\delta^T{\cal M} = \delta_i^S{\cal M} =0$.}
\begin{equation}\label{V.4}
    \delta^T \Phi(\omega, \mathbf{k}) = i \epsilon \omega \Phi(\omega, \mathbf{k}),
\end{equation}
where $\epsilon$ is an infinitesimal parameter. It follows that
\begin{equation}
\delta^T \Phi(S(\omega), S(\mathbf{k})) = i \epsilon S(\omega)
\Phi(S(\omega), S(\mathbf{k})) = -i \epsilon \omega
\Phi(S(\omega), S(\mathbf{k}))
\end{equation}
and using Leibniz rule we easily see
that the action is indeed invariant. Let us now consider the space
translation. Assume that in this case
\begin{equation}\label{V.5}
    \delta^S_i \Phi(\omega, \mathbf{k}) = i \epsilon k_i \Phi(\omega, \mathbf{k}).
\end{equation}
But then
\begin{equation}
\delta^S_i \Phi(S(\omega), S(\mathbf{k})) = i \epsilon S(k_i)\, \Phi(S(\omega), S(\mathbf{k})) = -i \epsilon e^\omega\,  k_i\, \Phi(S(\omega), S(\mathbf{k}))
\end{equation}
and the action is not invariant, if we apply Leibniz rule.
One can improve the situation by assuming that
$$\delta^S_i \Phi(\omega, \mathbf{k}) = i \epsilon e^{\omega/2}\, k_i
\Phi(\omega, \mathbf{k}),$$
which along with the appropriately
redefined transformation for $\Phi(S(\omega), S(\mathbf{k}))$ will
make the action invariant, but then such defined symmetries would
not form representation of the original $\kappa$-Poincar\'e
algebra.

The way out of this problem is to modify the Leibniz rule by the
co-product one (\ref{copro1}), as in \cite{Agostini:2003vg}. To
this end we take
$$
\delta^S_i\left( \Phi(\omega, \mathbf{k}) \Phi(S(\omega),
S(\mathbf{k})) \right)
    \equiv \delta^S_i\left( \Phi(\omega, \mathbf{k})\right) \Phi(S(\omega),
    S(\mathbf{k})) 
    + \left(e^{-\omega} \Phi(\omega, \mathbf{k})\right)\,
    \delta^S_i\left(\Phi(S(\omega), S(\mathbf{k})) \right) =0
$$
i.e., we generalize Leibniz rule by multiplying $\Phi(\omega, \mathbf{k})$
in the second term by $e^{-\omega}$. Note that such definition is consistent with
the fact that the fields are commuting, because
$$
  \delta^S_i\left( \Phi(S(\omega), S(\mathbf{k})) \Phi(\omega, \mathbf{k}) \right) = \left( i \epsilon\, S( k_i) + i \epsilon\, e^{-S(\omega)}\, k_i\right)\Phi(S(\omega), S(\mathbf{k})) \Phi(\omega, \mathbf{k}) =0.
$$
We see therefore that in order to make the action invariant with respect to infinitesimal phase transformations one must generalize the standard Leibniz rule to the non-symmetric co-product one.
This rule is well known from the theory of quantum groups and as shown in \cite{Agostini:2003vg} its emergence is a direct consequence of non-commutative structure of space-time. We see that the same quantum group structure arises also in momentum space picture.

Let us note at this point that the co-product rule cannot be generalized to a product of two arbitrary functions. For example
\begin{equation}\label{prod1}
\delta^S_i\left( f(\omega_1, \mathbf{k}_1)\, g(\omega_2, \mathbf{k}_2)\right) = \left(\mathbf{k}_1 + e^{-\omega_1}\, \mathbf{k}_2\right) f(\omega_1, \mathbf{k}_1)\, g(\omega_2, \mathbf{k}_2) \neq \delta^S_i\left(g(\omega_2, \mathbf{k}_2) \, f(\omega_1, \mathbf{k}_1)\right).
\end{equation}
It can be, in principle,  generalized to expressions of the form
$$
f \cdot g = \int\, d\omega_1\, d \omega_2\, d \mathbf{k}_1\, d \mathbf{k}_2\, {\cal K}(\omega_1, \omega_2; \mathbf{k}_1, \mathbf{k}_2)\, f(\omega_1, \mathbf{k}_1)\, g(\omega_2, \mathbf{k}_2)
$$
where the kernel ${\cal K}(\omega_1, \omega_2; \mathbf{k}_1, \mathbf{k}_2)$ is symmetric, because then only symmetrized part of $\delta^S_i$ is picked up. However such a generalization would be highly unnatural, because in the theory we started with we had to do only with integrals of products of functions taken at the same space-time point. It seems that without additional mathematics one cannot easily generalize the co-product rule beyond the products of the functions of the form
\begin{equation}\label{prod}
   f \star g = \int\, d\omega\,  d \mathbf{k}\, {\cal K}(\omega; \mathbf{k})\, f(\omega, \mathbf{k})\, g(S(\omega), S(\mathbf{k})).
\end{equation}
An immediate consequence of this is that, as mentioned above, it seems impossible within this formalism to define the energy and momentum of two particle states of the form $\hat a^\dag_{\mathbf{k}}\,\hat a^\dag_{\mathbf{k}'}\, |0>$.
\newline

Let us return to the boost transformations (\ref{V.1}). We saw above that the naive reasoning led us to problem: how to reconcile the transformation law (\ref{V.2}) with the rules (\ref{V.2a}), (\ref{V.2b}). As we know already the field $\Phi(\omega, \mathbf{k})$ must transform as a scalar, so that the action of all the symmetries on one field closes. We must therefore derive the  co-product, generalized Leibniz rule so that the product $\Phi(\omega, \mathbf{k}) \Phi(S(\omega), S(\mathbf{k}))$ transforms as a scalar as well. Thus instead of
$$
\delta^B\,\left({\Phi}(\omega, \mathbf{k})\, \Phi(S(\omega), S(\mathbf{k})) \right) \neq \delta^B\,\left({\Phi}(\omega, \mathbf{k})\right) \, \Phi(S(\omega), S(\mathbf{k})) + {\Phi}(\omega, \mathbf{k}) \, \delta^B \left(\Phi(S(\omega), S(\mathbf{k}))\right),
$$
where $\delta^B\,\left({\Phi}(\omega, \mathbf{k})\right)$ and $\delta^B\,\left(\Phi(S(\omega), S(\mathbf{k}))\right)$ are given by (\ref{V.2a}) and (\ref{V.2b}), respectively, we should consider its appropriate generalization.

As a first step, observe that since $\omega =-S(\omega)$, $k_i = - e^{S(\omega)}\, S(k_i)$,
$$
\frac{\partial}{\partial S(\omega)} = - \frac{\partial}{\partial \omega} + k_i\, \frac{\partial}{\partial k_i}, \quad \frac{\partial}{\partial S(k_i)} =  - e^{-\omega}\, \frac{\partial}{\partial k_i}
$$
and
\begin{equation}\label{V.9}
 \delta^B\,{\Phi}(S(\omega), S(\mathbf{k})) = \left[\varepsilon^i\, e^{\omega}\, k_i\,  \frac{\partial}{\partial \omega}  + \left({1\over 2}\varepsilon_i \left(
 e^{\omega} -e^{-\omega}
 - e^{\omega}\,\mathbf{k}^2)  \right) \right)  \frac{\partial}{\partial k_i}\right] {\Phi}(S(\omega), S(\mathbf{k}))
\end{equation}
Therefore
\begin{eqnarray}
\delta^B\left({\Phi}(\omega, \mathbf{k})\, {\Phi}(S(\omega),
S(\mathbf{k}))\right) &=& \delta^B\left\{{\Phi}(\omega,
\mathbf{k})\right\}\, {\Phi}(S(\omega), S(\mathbf{k})) \nonumber
\\
&+& \left\{e^{-\omega}\, {\Phi}(\omega, \mathbf{k})\right\}\,
\delta^B\left\{ {\Phi}(S(\omega), S(\mathbf{k}))\right\} \nonumber
\\
&+&
   {\Phi}(\omega, \mathbf{k})\, \left[-\varepsilon^i\,  k_i\, k_j\,
   \frac{\partial}{\partial k_j} + \varepsilon^i\, \mathbf{k}^2\,
   \frac{\partial}{\partial k_i} \right] {\Phi}(S(\omega), S(\mathbf{k}))
\end{eqnarray}
This last term can be expressed  as
\begin{equation}
\varepsilon^i\, \epsilon_{ijk} \, \left\{ k_j \, {\Phi}(\omega, \mathbf{k})\right\} \, \left\{ \epsilon_{klm} \, k_l\, \frac{\partial}{\partial k_m}\, {\Phi}(S(\omega), S(\mathbf{k}))\right\}.
\end{equation}
Observe however that
\begin{equation}
\epsilon_{klm} \, k_l\, \frac{\partial}{\partial k_m} = \epsilon_{klm} \, S(k_l)\, \frac{\partial}{\partial S(k_m)}
\end{equation}
 is nothing but the rotation generator $M_k$ so what we have is
just the co-product deformed Leibniz rule for boosts (\ref{copro2}).
\newline

Let us summarize. We have shown that the momentum  space action (\ref{29}) is invariant under infinitesimal transformations of the fields of $\delta^T$, $\delta_i^S$, and $\delta^B$ given by (\ref{V.4}), (\ref{V.5}), and (\ref{V.2a}), respectively. It is clear that commutators of these transformations form a ten parameter algebra of symmetries, which is isomorphic to the algebraic part of the original $\kappa$-Poincar\'e algebra (\ref{1})--(\ref{5}). Moreover, the action of these symmetries on products of functions must be   given by co-product rule (\ref{copro1}), (\ref{copro2}). This proves that our action is indeed invariant under ten parameter, quantum $\kappa$-Poincar\'e algebra.

\section{Interactions and conservation laws}

Till now we have been considering free theory, however from the point of view of DSR phenomenology \cite{Amelino-Camelia:2002vw}, \cite{Amelino-Camelia:2003ex} it is the interacting theory that is of real interest. Here we will concentrate on the $\Phi^4(x)$ interaction, which is of major interest in view of phenomenological application, like possible explanation of  violation of GZK cut-off and TeV photons anomaly (see, e.g., \cite{Kifune:1999ex}, \cite{Protheroe:2000hp}.)

Let us start with the natural form of the interaction term
\begin{equation}\label{VI.1}
   S^{int} = \lambda \int dt d^3\mathbf{x}\, \Phi^4(t, \mathbf{x})
\end{equation}
Going to  Fourier transformed fields we find
\begin{eqnarray}
S^{int} &=& \lambda \int d\omega_1 d^3 \mathbf{k}_1\, d\omega_2
d^3 \mathbf{k}_2\, d\omega_3 d^3 \mathbf{k}_3\, d\omega_4 d^3
\mathbf{k}_4\, \Phi(\omega_1, \mathbf{k}_1)\Phi(\omega_2,
\mathbf{k}_2)\Phi(\omega_3, \mathbf{k}_3)\Phi(\omega_4,
\mathbf{k}_4)\,\times \nonumber \\
&\times& \,
\delta\left(\omega_1+\omega_2+\omega_3+\omega_4\right)\,\delta\left(e^{-\omega_2-
\omega_3- \omega_4}\,\mathbf{k}_1 + e^{-\omega_3- \omega_4}\,
\mathbf{k}_2 + e^{-\omega_4}\, \mathbf{k}_3+
\mathbf{k}_4\right)\label{VI.2}
\end{eqnarray}
Note that as a result of the presence of
$ \delta\left(\omega_1+\omega_2+\omega_3+\omega_4\right)$ we could have
dropped the term $e^{\omega_1+\omega_2+\omega_3+\omega_4}$ in the lagrangian.

Formula (\ref{VI.2}) looks  not symmetric
with respect to exchange of labels $1,\ldots,4$. However
the expression in delta function is symmetrized because the
fields $\Phi(\omega_i, \mathbf{k}_i)$, $i=1,\ldots,4$ commute.
The action (\ref{VI.2}) can be therefore rewritten as
\begin{eqnarray}
S^{int} &=& \lambda \int d\omega_1 d^3 \mathbf{k}_1\, d\omega_2
d^3 \mathbf{k}_2\, d\omega_3 d^3 \mathbf{k}_3\, d\omega_4 d^3
\mathbf{k}_4\,\times \nonumber \\
&\times & \Phi(\omega_1, \mathbf{k}_1)\Phi(\omega_2,
\mathbf{k}_2)\Phi(\omega_3, \mathbf{k}_3)\Phi(\omega_4,
\mathbf{k}_4)\,
\delta\left(\omega_1+\omega_2+\omega_3+\omega_4\right)\, \times
\nonumber \\
&\times &\frac{1}{4!}\, \sum_{ i(j)}\delta\left(e^{-\omega_{i(2)}-
\omega_{i(3)}- \omega_{i(4)}}\,\mathbf{k}_{i(1)} +
e^{-\omega_{i(3)}- \omega_{i(4)}}\, \mathbf{k}_{i(2)} +
e^{-\omega_{i(4)}}\, \mathbf{k}_{i(3)}+  \mathbf{k}_{i(4)}\right)
\label{VI.3}
\end{eqnarray}

This result is in direct analogy to quantum mechanical computations presented in \cite{Amelino-Camelia:2001fd}. In this paper the authors found, by using the path integral quantization, that the conservation rule associated with four-point vertex in $\kappa$-Minkowski field theory has the form of 24 delta functions, exactly as in (\ref{VI.3}). Our interpretation of (\ref{VI.3}) derives from the one of \cite{Amelino-Camelia:2001fd}: the theory  predicts that one
of the 24 conservation rules must be satisfied in a process classically, and assigns equal (quantum)
probabilities to each of these 24 channels.

Observe that every delta carries information about conservation law in the case of interactions of four identical particles, of the form
\begin{equation}\label{VI.4}
    \omega_1 + \omega_2 + \omega_3 + \omega_4 =0,
\end{equation}
where, of course the energies of incoming and outgoing particles should be taken with plus and minus signs, respectively, and
\begin{equation}\label{VI.5}
    e^{-\omega_2- \omega_3- \omega_4}\,\mathbf{k}_1 + e^{-\omega_3- \omega_4}\, \mathbf{k}_2 + e^{-\omega_4}\, \mathbf{k}_3+  \mathbf{k}_4 =0
\end{equation}
plus all 23 other permutations, of which only 6 are independent because of the invariance under cyclic permutations $1\rightarrow2\rightarrow3\rightarrow4\rightarrow1$.

In the standard case we have to do with 24 deltas as well, however they are all equal, and thus they lead to a single conservation law. Here each of the 6 independent deltas describes different momentum conservation law. This fact can be interpreted as a signal that the $\Phi^4(x)$ vertex in space-time turns into 6 different ways that momentum can be conserved. This is certainly odd, however does not seem to contradict any basic physical principle. One could interpret this as signal that in the case of a theory defined on $\kappa$-Minkowski space-time,  interacting particles can choose between different modes of energy momentum conservation, in other words, the theory local in  $\kappa$-Minkowski predicts 6 independent identical particle scattering processes. Of course, in the case of scattering of different particles, the number of independent processes would be larger, since the number of available permutations will be lower.

From the point of view of DSR phenomenology this would presumably
mean that given the exact energies and momenta of incoming
particles, and of one outgoing particle, we could not make any
predictions for energy and momentum of the second outgoing one.
Instead we would have some number of possibilities (6 in the case
of identical particles,) as to which its energy and momentum could
be. Of course this indeterminism disappears in the limit when
energies of all particles are small as compared to 1 (i.e., to
Planck scale.)

\section{Conclusions and open problems}

In this paper we wanted to emphasize that it is possible to discuss scalar field theory on $\kappa$-Minkowski space-time solely in terms of the momentum picture. In particular we show that the momentum picture action is invariant under the same ten-parameter (quantum) $\kappa$-Poincar\'e algebra of symmetries as the original action defined on $\kappa$-Minkowski space-time. We also proposed to make use of the isomorphism between complex and symplectic structure in order to quantize the theory. Finally we analyzed $\Phi^4(x)$ interaction term that can be added to free field action, and we show that, in analogy to the results of \cite{Amelino-Camelia:2001fd}, it leads to a theory possessing 6 independent energy-momentum conservation rules, which presumably leads to classical indeterminism of the outcome of  scattering process.

Of course a great number of open problems remains. In our view the one concerning construction of Fock space seems to be the most urgent one. Namely, if one uses the construction of Sect.\ IV and identifies the field operators ${\Phi}(\omega(\mathbf{k}),  \mathbf{k})$ and  $  {\Phi}^*(\omega(\mathbf{k}),  \mathbf{k})$,  $\omega(\mathbf{k})>0$ with annihilation and creation operators, respectively,  one cannot define energy and momentum of many-particles DSR states constructed in the standard way. This is because the co-product action is not defined for ordinary products of functions (\ref{prod1}). It seems inevitable that the definition of many-particles states must differ from the standard case, so as to ensure mathematical consistency and, moreover, to lead to physically acceptable results. For example it would be clearly not acceptable if an expression for total energy and momentum of a system of identical particles is not symmetric in permutation of the particles. We will address these problems in a future publication.

\section*{Acknowledgement}
We would like to thank Carlo Rovelli for discussion and bringing ref.\ \cite{Rovelli:1999fq} to our attention.


\begin{thebibliography}{99}
\bibitem{kappaM} S.~Majid and H.~Ruegg, ``Bicrossproduct structure of kappa Poincare group
and noncommutative geometry,'' Phys.\ Lett.\ B {\bf 334} (1994) 348
[arXiv:hep-th/9405107]; J.~Lukierski, H.~Ruegg and W.~J.~Zakrzewski,
``Classical quantum mechanics of free kappa relativistic systems,'' Annals
Phys.\  {\bf 243} (1995) 90 [arXiv:hep-th/9312153].

\bibitem{Kowalski-Glikman:2002jr}
J.~Kowalski-Glikman and S.~Nowak, ``Non-commutative space-time of doubly
special relativity theories,'' Int.\ J.\ Mod.\ Phys.\ D {\bf 12} (2003) 299
[arXiv:hep-th/0204245].

\bibitem{Kowalski-Glikman:2002ft}
J.~Kowalski-Glikman, ``De Sitter space as an arena for doubly special
relativity,'' Phys.\ Lett.\ B {\bf 547} (2002) 291 [arXiv:hep-th/0207279].



\bibitem{Amelino-Camelia:2000ge}
G.~Amelino-Camelia, ``Testable scenario for relativity with minimum-length,''
Phys.\ Lett.\ B {\bf 510}, 255 (2001) [arXiv:hep-th/0012238].

\bibitem{Amelino-Camelia:2000mn}
G.~Amelino-Camelia, ``Relativity in space-times with short-distance structure
governed by an observer-independent (Planckian) length scale,'' Int.\ J.\ Mod.\
Phys.\ D {\bf 11}, 35 (2002) [arXiv:gr-qc/0012051].
\bibitem{jkgminl} J.~Kowalski-Glikman,
``Observer independent quantum of mass,'' Phys.\ Lett.\ A {\bf 286} (2001) 391
[arXiv:hep-th/0102098].

\bibitem{rbgacjkg} N.~R.~Bruno, G.~Amelino-Camelia and J.~Kowalski-Glikman,
``Deformed boost transformations that saturate at the Planck scale,'' Phys.\
Lett.\ B {\bf 522} (2001) 133 [arXiv:hep-th/0107039].

\bibitem{Amelino-Camelia:2003xp}
G.~Amelino-Camelia, L.~Smolin and A.~Starodubtsev,
``Quantum symmetry, the cosmological constant and Planck scale
Class.\ Quant.\ Grav.\  {\bf 21} (2004) 3095
[arXiv:hep-th/0306134].



\bibitem{Freidel:2003sp}
L.~Freidel, J.~Kowalski-Glikman and L.~Smolin,
``2+1 gravity and doubly special relativity,''
Phys.\ Rev.\ D {\bf 69} (2004) 044001
[arXiv:hep-th/0307085].



\bibitem{Kowalski-Glikman:2003hi}
J.~Kowalski-Glikman,
``Doubly special relativity and quantum gravity phenomenology,''
arXiv:hep-th/0312140, Proceedings of 10th Marcel Grossmann Meeting, to appear; J.~Kowalski-Glikman,
``Introduction to doubly special relativity,''
arXiv:hep-th/0405273, Lecture Notes in Physics, to appear.


\bibitem{Amelino-Camelia:2002vy}
G.~Amelino-Camelia, ``Doubly-Special Relativity: First Results and Key Open
Problems,'' Int.\ J.\ Mod.\ Phys.\ D {\bf 11} (2002) 1643
[arXiv:gr-qc/0210063];
G.~Amelino-Camelia,
 ``Some encouraging and some cautionary remarks on doubly special relativity in
quantum gravity,''
arXiv:gr-qc/0402092.



\bibitem{juse} J.~Kowalski-Glikman and S.~Nowak,
``Doubly special relativity theories as different bases of kappa-Poincare
algebra,'' Phys.\ Lett.\ B {\bf 539} (2002) 126 [arXiv:hep-th/0203040].

\bibitem{Magueijo:2001cr}
J.~Magueijo and L.~Smolin, ``Lorentz invariance with an invariant energy
scale,'' Phys.\ Rev.\ Lett.\  {\bf 88} (2002) 190403 [arXiv:hep-th/0112090].

\bibitem{qP} J.~Lukierski, H.~Ruegg, A.~Nowicki and V.~N.~Tolstoi,
``Q deformation of Poincar\'e algebra,'' Phys.\ Lett.\ B {\bf 264}
(1991) 331; J.~Lukierski, A.~Nowicki and H.~Ruegg, ``New quantum
Poincare algebra and k deformed field theory,'' Phys.\ Lett.\ B
{\bf 293} (1992) 344.

\bibitem{Kosinski:1999ix}
P.~Kosinski, J.~Lukierski and P.~Maslanka, ``Local D = 4 field theory on
kappa-deformed Minkowski space,'' Phys.\ Rev.\ D {\bf 62} (2000) 025004
[arXiv:hep-th/9902037].

\bibitem{Kosinski:2001ii}
P.~Kosinski, J.~Lukierski and P.~Maslanka,
 ``kappa-deformed Wigner construction of relativistic wave functions and  free
fields on kappa-Minkowski space,''
Nucl.\ Phys.\ Proc.\ Suppl.\  {\bf 102} (2001) 161
[arXiv:hep-th/0103127].




\bibitem{Kosinski:2003xx}
P.~Kosinski, P.~Maslanka, J.~Lukierski and A.~Sitarz,
 ``Generalized kappa-deformations and deformed relativistic scalar fields on
noncommutative Minkowski space,''
arXiv:hep-th/0307038.

\bibitem{Amelino-Camelia:2001fd}
G.~Amelino-Camelia and M.~Arzano, ``Coproduct and star product in field
theories on Lie-algebra  non-commutative space-times,'' Phys.\ Rev.\ D {\bf 65}
(2002) 084044 [arXiv:hep-th/0105120].

\bibitem{Dimitrijevic:2004nv}
M.~Dimitrijevic, L.~Jonke, L.~Moller, E.~Tsouchnika, J.~Wess and M.~Wohlgenannt,
``Field theory on kappa-spacetime,''
arXiv:hep-th/0407187;
M.~Dimitrijevic, F.~Meyer, L.~Moller and J.~Wess,
``Gauge theories on the kappa-Minkowski spacetime,''
Eur.\ Phys.\ J.\ C {\bf 36} (2004) 117
[arXiv:hep-th/0310116];
M.~Dimitrijevic, L.~Jonke, L.~Moller, E.~Tsouchnika, J.~Wess and M.~Wohlgenannt,
``Deformed field theory on kappa-spacetime,''
Eur.\ Phys.\ J.\ C {\bf 31} (2003) 129
[arXiv:hep-th/0307149].

\bibitem{Agostini:2003vg}
A.~Agostini, G.~Amelino-Camelia and F.~D'Andrea,
``Hopf-algebra description of noncommutative-spacetime symmetries,''
arXiv:hep-th/0306013.

\bibitem{5dcalc1} A.~Sitarz, ``Noncommutative differential calculus on the kappa
Minkowski space,'' Phys.\ Lett.\ B {\bf 349} (1995) 42 [arXiv:hep-th/9409014].

\bibitem{5dcalc2} J.A.~de~Azcarraga and J.C.~P\'erez Bueno, ``Relativistic and Newtonian
$\kappa$-spacetimes'', J.~Math.~Phys. {\bf 36} (1995) 6879
[arXive:q-alg/9505004]; P.~Kosinski, P.~Maslanka, and J.~Sobczyk, ``The
bicovariant differential calculus on the $\kappa$-Poincar\'e group and on the
$\kappa$-Minkowski space'' [arXive:q-alg/9508021].

\bibitem{5dcalc3} S.~Giller, C.~Gonera,
P.~Kosinski, and P.~Maslanka, ``A note on geometry of $\kappa$-Minkowski
space'', Acta Phys.\ Polon.\ B {\bf 27} (1996) 2171 [arXive:q-alg/9602006];
P.~Kosinski, and P.~Maslanka, J.~Lukierski, and A.~Sitarz, ``Towards
$\kappa$-deformed D=4 relativistic field theory'', Chech.\ J.\ Phys.\ {\bf 48}
(1998) 1407.

\bibitem{Kowalski-Glikman:2003we}
J.~Kowalski-Glikman and S.~Nowak, ``Doubly special relativity and de Sitter
space,'' Class.\ Quant.\ Grav.\  {\bf 20} (2003) 4799
[arXiv:hep-th/0304101].


\bibitem{Kowalski-Glikman:2001ct}
J.~Kowalski-Glikman,
 ``Doubly special quantum and statistical mechanics from quantum
$\kappa$-Poincar\'e algebra,''
Phys.\ Lett.\ A {\bf 299} (2002) 454
[arXiv:hep-th/0111110].

\bibitem{Blaut:2001fy}
A.~Blaut, J.~Kowalski-Glikman and D.~Nowak-Szczepaniak,
``kappa-Poincare dispersion relations and the black hole radiation,''
Phys.\ Lett.\ B {\bf 521} (2001) 364
[arXiv:gr-qc/0108069]; J.~Kowalski-Glikman,
``Testing dispersion relations of quantum kappa-Poincare algebra on
cosmological ground,''
Phys.\ Lett.\ B {\bf 499} (2001) 1
[arXiv:astro-ph/0006250].







%
\bibitem{Amelino-Camelia:2002vw}
G.~Amelino-Camelia,
``Quantum-gravity phenomenology: Status and prospects,''
Mod.\ Phys.\ Lett.\ A {\bf 17} (2002) 899
[arXiv:gr-qc/0204051].




\bibitem{Amelino-Camelia:2003ex} 
G.~Amelino-Camelia, J.~Kowalski-Glikman, G.~Mandanici and A.~Procaccini,
``Phenomenology of doubly special relativity,''
arXiv:gr-qc/0312124.

\bibitem{Rovelli:1999fq}
C.~Rovelli,
``Spectral noncommutative geometry and quantization: A simple example,''
Phys.\ Rev.\ Lett.\  {\bf 83} (1999) 1079
[arXiv:gr-qc/9904029].

\bibitem{Kifune:1999ex}
T.~Kifune,
``Invariance violation extends the cosmic ray horizon?,''
Astrophys.\ J.\  {\bf 518} (1999) L21
[arXiv:astro-ph/9904164].

\bibitem{Protheroe:2000hp}
R.~J.~Protheroe and H.~Meyer,
``An infrared background TeV gamma ray crisis?,''
Phys.\ Lett.\ B {\bf 493} (2000) 1
[arXiv:astro-ph/0005349].






\end{thebibliography}
\end{document}